\documentclass[conference]{IEEEtran}
\usepackage[utf8]{inputenc}
\usepackage{cite}
\usepackage{amsmath,amssymb,amsfonts, amsthm}
\usepackage{algorithmic}
\usepackage{graphicx}
\usepackage{textcomp}
\usepackage[caption=false]{subfig}
\usepackage[justification=centering]{caption}
\usepackage[ruled,vlined]{algorithm2e}
\usepackage{epsfig}
\usepackage{float}
\usepackage{color}
\usepackage{balance}
\usepackage{bbm}
\usepackage{textcomp}
\usepackage{enumitem}
\usepackage{url}
\usepackage{comment}
\usepackage[normalem]{ulem}
\usepackage[flushleft]{threeparttable}
\usepackage{footnote}

\newtheorem{remark}{Remark}

\newtheorem{lemma}{Lemma}

\newcommand\blfootnote[1]{%
  \begingroup
  \renewcommand\thefootnote{}\footnote{#1}%
  \addtocounter{footnote}{-1}%
  \endgroup
}

\newcommand{\fig}[1]{Fig. \ref{#1}}

\newcommand{\argmax}{\mathrm{arg\,max}}
\newcommand{\argmin}{\mathrm{arg\,min}}
\newcommand{\mathrmbold}[1]{\boldsymbol{\mathrm{#1}}}

\newcommand{\wrt}{\textit{w.r.t.~}}

\newcommand{\ien}{\textit{i.e.}}

\newcommand{\egn}{\textit{e.g.}}

\newcommand{\norm}[1]{\lVert#1\rVert}

\def\xweights{\mathrmbold{\theta}}
\def\yweights{\mathrmbold{\vartheta}}

\def\mb{\mathrmbold{m}}

\def\zb{\mathrmbold{z}}
\def\zpb{\mathrmbold{z}'}
\def\yb{\mathrmbold{y}}
\def\xb{\mathrmbold{x}}
\def\dx{\mathrm{d}x}
\def\dy{\mathrm{d}y}

\title{Learning Semantics: An Opportunity for Effective 6G Communications}
\author{\IEEEauthorblockN{Mohamed Sana$^{1}$, Emilio Calvanese Strinati$^{1}$}

\IEEEauthorblockA{$^{1}$CEA-Leti, Université Grenoble Alpes, F-38000 Grenoble, France\\
Email : \{mohamed.sana, emilio.calvanese-strinati\}@cea.fr}}
\date{April 2021}

\begin{document}
\maketitle\blfootnote{This work was partly funded by the H2020 project Hexa-X (Grant Agreement no. 101015956).}  
\maketitle
\begin{abstract}
Recently, semantic communications are envisioned as a key enabler of future 6G networks. Back to Shannon's information theory, the goal of communication has long been to guarantee the correct reception of transmitted messages irrespective of their meaning. However, in general, whenever communication occurs to convey a meaning, what matters is the receiver's understanding of the transmitted message and not necessarily its correct reconstruction. Hence, semantic communications introduce a new paradigm: transmitting only relevant information sufficient for the receiver to capture the meaning intended can save significant communication bandwidth. Thus, this work explores the opportunity offered by semantic communications for beyond 5G networks. In particular, we focus on the benefit of semantic compression. We refer to \textit{semantic} message as a sequence of well-formed symbols learned from the ``meaning'' underlying data, which have to be interpreted at the receiver. This requires a reasoning unit, here artificial, on a knowledge base: a symbolic knowledge representation of the specific application. Therefore, we present and detail a novel architecture that enables \emph{representation learning} of semantic symbols for effective semantic communications. We first discuss theoretical aspects and successfully design objective functions, which help learn effective semantic encoders and decoders. Eventually, we show promising numerical results for the scenario of text transmission, especially when the sender and receiver speak different languages.
\end{abstract}

\section{Introduction}
\noindent
Academia and industry have kicked off research on the future sixth generation (6G) of wireless networks. Speculation about the possible evolution of current 5G technology as well as radical new architectures, approaches, and technologies are being intensely discussed \cite{Calvanese6GVTM2019, Chowdhury2020}. This is driven by the current trend, witnessing an unprecedented demand for communication bandwidth to accommodate burgeoning new services like extended reality or autonomous driving. To meet these challenges, historically in wireless communications, a solution has been to explore higher frequencies to benefit from the available large spectrum resources. Such solutions cyclically face an inevitable bottleneck, represented by the hardware's cost, the complexity and energy efficiency of wireless communications. For example, as frequency increases new challenges arise in communication such as blockage, severe pathloss, atmospheric absorption, and power amplifier efficiency \cite{belot2020spectrum}. This calls for new paradigms shift for the effective design of 6G communications \cite{CalvaneseGOWSC2021}. In our view, future 6G systems should be engineered to effectively recreate or infer the meaning of what has been communicated rather than to \textit{``just''} optimize opaque data pipes aiming to reproduce exactly exchanged sequences of symbols \cite{CalvaneseGOWSC2021}. Effective communication of \textit{meanings} can be achieved through exchanges of \textit{semantics}. However, \emph{how to bring the notion of \emph{semantic} from human understanding to machine understanding?} In our view, this requires a radically innovative approach to communications: the semantic and goal-oriented communications \cite{CalvaneseGOWSC2021}. This approach can achieve a significant source data compression gain, which saves a lot of communication bandwidth.

In their seminal work \cite{shannon1948mathematical, weaver1953recent}, Shannon and Weaver identified three levels of communication:
\begin{itemize}
    \item \emph{Level A} - the technical problem: how accurately can the symbols of communication be transmitted?
    \item \emph{Level B} - the semantic problem: how precisely do the transmitted symbols convey the desired meaning?
    \item \emph{Level C} - the effectiveness problem: how effectively does the received meaning affect conduct in the desired way?
\end{itemize}

Shannon deliberately focused on the technical problem and the communication systems that we know so far are engineered to optimize the \emph{level A} of communication. Then in 1953, Weaver provided a first attempt for the inclusion of semantics in the communication problem \cite{weaver1953recent}. Bar-Hillel and Carnap provided also outlines of a theory of semantic information, focusing mainly on measuring how informative transmitted message is (informativity measurement) \cite{BarHillel1964}. Recently with the growing interdependence between communication systems and artificial intelligence (AI), new attempts to include the \emph{Level B} in the communication system has started \cite{CalvaneseGOWSC2021, Bao2011, Huiqiang2021a, Shi2021}. 
%

For \cite{CalvaneseGOWSC2021}, semantic communications must be shaped to effectively compress the exchanged data between communicating parties, improve the communication robustness by incorporating semantic information to the classical \textit{Level A} communication scheme. This is possible by exploiting the knowledge shared a priori between communicating parties, such as a shared language or logic, shared background and contextual knowledge, and possibly a shared view on the goal of communication. In \cite{Bao2011}, the authors provide tentative definitions of semantic capacity, semantic noise, and a semantic channel from the perspective of Shannon's statistical measurement of information. In \cite{kountouris2020}, the authors refers to semantic as the \textit{semantics of information}, addressing the significance and usefulness of messages by considering the contextual attributes (semantics) of information \cite{KostaAgeoI2017}. In this approach, the age of information (AoI) is key to identify the relevance of the semantic information for the effectiveness of the exchange between communicating parties. Nevertheless, AoI does not necessarily define the meaning of a message in many applications, but rather how a message is still pertinent for an application given its age. 
In \cite{Huiqiang2021a}, an end-to-end (E2E) neural architecture is presented enabling semantic transmission of sentences. However, their proposed architecture is limited in flexibility: they represent each word in a transmitted sentence with the same and fixed number of semantic symbols irrespective of the conveyed meaning. Authors in \cite{weng2020semantic} apply the same architecture to speech signals transmission. Similarly, the work in \cite{jiang2021deep} presents a deep source-channel coding scheme, which exploits hybrid automatic repeat request (HARQ) to reduce semantic transmission error.
Here in this work, we focus on the benefit of semantic compression. We refer to \textit{semantic} as a ``meaningful'' message (a sequence of well-formed symbols, which are possibly learned from data) that have to be interpreted at the receiver. This requires a reasoning unit (natural or artificial) able to interpret based on a knowledge base: a symbolic knowledge representation of the specific application. 

Following this approach, we propose a novel E2E semantic communication architecture incorporating level B to classical level A communications. In this architecture, information from a binary source is encoded with semantic information extracted using neural attention mechanisms \cite{vaswani2017attention}, to produce a sequence of semantic symbols. In contrast to very recent state-of-the-art works \cite{Huiqiang2021a, weng2020semantic}, which propose an E2E system for semantic text and speech transmission, we formally define a new loss function, which captures the effects of \emph{semantic distortion} to communication. This enables to dynamically trade semantic compression losses with semantic fidelity \cite{Bao2011} (\ien, the semantic interpretation correctness). In addition, we design a semantic adaptive mechanism, which dynamically optimizes the number of symbols per semantic message based on the trade-off between semantic compression and semantic fidelity that we formally express. Eventually, we provide a detailed numerical evaluation that shows the benefits of our proposed adaptive E2E semantic system. Results are provided for the context of natural language processing (NLP), especially when the sender and receiver speak a different language. In this context, messages are formed and communication parameters are set to maximize the correct interpretation of semantic messages rather than error-free bit decoding at the receiver.

\section{Semantic communications: system model}

A semantic communication system defines a communication framework in which the sender and receiver exchange \emph{semantic information} to create a common understanding of exchanged messages. We refer to the term \emph{semantic information} as the meaning underlying the data (which can be discrete or continuous) that a sender wants to convey to a receiver. Example of data ranges from (random) numbers to texts, audios, images or videos. 
Here, we focus on applications where AI agents exchange, communicate and intertwine. For this, we adopt \emph{semantic symbols} as a means to represent semantics. In contrast to classical source-channel coding schemes aiming to optimize channel modulation with opaque data pipes (sequence of bits), here in our scenario, agents exchange semantic symbols depending on the meaning associated with the exchanged data. These symbols are generated at the transmitter using a semantic encoder $f_{\xweights}(\cdot)$, and interpreted at the receiver using a semantic decoder $g_{\yweights}(\cdot)$ after a proper design of a \emph{semantic representation learning} loss, which we describe below. 


\subsection{Semantic source and channel coding}
The semantic encoder transforms the input sequence into semantic symbols to be transmitted through the channel. Let $\mathcal{M}_t$ denotes the source alphabet. Each message $m$ emitted by the source is associated with a symbol $x\in\mathcal{X}$ (possibly a discrete or continuous space) such that $x = f_{\xweights}(m)$, where $f_{\xweights}(\cdot)$ denotes the semantic encoder with (trainable) parameters $\xweights$. This encoder is characterized by the probability distribution $p_{\xweights}(x|m)$. Thus, if the source emits a message $m$ with a probability $p_{\mathcal{M}_t}(m)$, the probability that the encoder emits symbol $x$ is $p_{\xweights}(x) = \sum_{\substack{m: x=f_{\xweights}(m)\\m\in\mathcal{M}_t}} p_{\mathcal{M}_t}(m)$, which gives:
\begin{align}\label{eq:probx}
    p_{\xweights}(x) \overset{\Delta}{=}\sum_{m\in\mathcal{M}_t} \delta\left(x - f_{\xweights}(m)\right) p_{\mathcal{M}_t}(m),
\end{align}
where $\delta(\cdot)$ is the Dirac distribution. Next, our objective is to define the adequate symbols probability distribution $p_{\xweights}(x)$ (or equivalently $p_{\xweights}(x|m)$), which ensures semantic ``fidelity" of interpreted messages at the receiver. Note that the mapping from $m$ to $x$ is not always bijective \cite{BASU2014188}. Indeed, it can be one-to-many: a message can be mapped to different symbols, each conveying the same information. In this case, the encoder introduces \emph{semantic redundancy}, \ien, the conditioned entropy $H_{\xweights}(X|M) \neq 0$. Conversely, the mapping can be many-to-one, \ien, many messages are mapped to the same symbol: there is a \emph{semantic ambiguity}, and $H_{\xweights}(M|X) \neq 0$. As in the rate-distortion Theory \cite{shannon1959coding}, such an encoder has a complexity equals to $I_{\xweights}(X;M)$, which corresponds to the average number of bits needed to represent message $m$. Hence, as we focus on semantic compression, our first objective is to find $f_{\xweights}(\cdot)$, which minimizes this complexity \ien,
\begin{align}\label{eq:cpobj}
    \underset{\xweights}{\argmin}~I_{\xweights}(X;M)
\end{align}

\begin{lemma}
    If there is no redundancy introduced by the semantic encoder, \ien, $M$ determines $X$ as the mapping $f_{\xweights}:~\mathcal{M}_t\rightarrow\mathcal{X}$ is unique, then, 
    \begin{align}
        I_{\xweights}(X;M) = H_{\xweights}(X).
    \end{align}
    In this case, minimizing $I_{\xweights}(X;M)$ or $H_{\xweights}(X)$ are equivalent.
    \begin{proof}
        First note $I_{\xweights}(X;M) = H(X) - H_{\xweights}(X|M)$. Thus, proof follows as $H_{\xweights}(X|M)=0$ if there is no redundancy.
    \end{proof}
\end{lemma}

\subsection{Semantic decoder}
The role of the decoder is mainly to infer the meaning (or an equivalent) intended by the source. In contrast to Shannon's communications paradigm, an exact reconstruction of the transmitted messages is not necessary. Given the receiver alphabet $\mathcal{M}_r$  and the semantic decoder $g_{\yweights}(\cdot)$ with (trainable) parameters $\yweights$, the decoded message $\hat{m}$ from a received symbol $y$ is the one that maximizes the estimated posterior probability $q_{\yweights}(m|y)$ conditioned on the received symbol $y$ at the receiver:
\begin{align}
    \hat{m} = \underset{m' \in \mathcal{M}_r}{\argmax~} q_{\yweights}(m'|y), 
\end{align}
Hence, given the semantic encoder and decoder, a natural measure of the semantic distortion between $m$ and $\hat{m}$ is the expected Kullback Leibler (KL) divergence between the "true" posterior probability $p_{\xweights}(m|y)$ at the encoder and the one captured by the decoder $q_{\yweights}(m|y)$, denoted $\mathrm{EKL}_{\xweights, \yweights}$ is
\begin{align}\label{eq:klfunc}
    \mathrm{EKL}_{\xweights, \yweights} &=\mathbb{E}_{y}\left\{\mathrm{KL}\left(p_{\xweights}(m|y)|| q_{\yweights}(m|y)\right)\right\} \\\nonumber
    &=\sum_{m\in\mathcal{M}_r} \int_y p_{\xweights}(y) p_{\xweights}(m|y) \mathrm{log}\left(\frac{q_{\yweights}(m|y)}{p_{\xweights}(m|y)} \right) \dy.
\end{align}
Our second objective is then to find $f_{\xweights}(\cdot)$ and $g_{\yweights}(\cdot)$ which minimize the semantic distortion between the intended message $m$ and the decoded message $\hat{m}$, \ien, Eqn. \eqref{eq:klfunc}.
\begin{align} \label{eq:klobj}
    \underset{\xweights, \yweights}{\argmin}~\mathrm{EKL}_{\xweights, \yweights}
\end{align}

\subsection{Semantic channel and noise}
A semantic channel defines a group of truth functions. For \egn, considers that ``Linda is in the kitchen'' and Simon is asking Michael to know where Linda is. If Michael answers ``Linda is cooking", then Simon can accurately interpret this answer as ``Linda is in the kitchen'': there is no \emph{semantic} failure. However, comparing these two sentences character by character may result in a classical Level A's communication failure. In this example, Michael serves as a \emph{semantic channel}. 

We say that two messages are semantically equivalent if they convey the same meaning. In other words, the received message $\hat{m}$ and the transmitted message $m$ are semantically equivalent if $\hat{m}$ is interpreted accurately by the receiver as the meaning intended by the transmitter. A formal definition of semantic equivalence is not trivial, as it can take different forms depending on the purpose of the communication and the type of data manipulated by the source and the destination. For example, in NLP, two words may be semantically equivalent if they are synonyms. A semantic error may occur during communication as the result of a mismatch between $m$ and $\hat{m}$: the two messages are not semantically equivalent. This error can be introduced by \textit{Level A} channel noise and/or interference, the difference of the level of knowledge available at the source and destination or its incompleteness (at \textit{Level B}) and, by limitation of semantic encoder/decoder not being able to learn the correct semantic representation, \ien, a limitation of the representation space of $f_{\xweights}(\cdot)$ and $g_{\yweights}(\cdot)$. To design an efficient communication system, given the semantic channel with probability density $p(y|x)$, our proposed solution maximizes the mutual information $I_{\xweights}(X;Y)$ between the input and output of the channel:
\begin{align}\label{eq:chobj}
    \underset{\xweights}{\argmax}~I_{\xweights}(X;Y)
\end{align}

\subsection{Proposed semantic representation learning}

To optimize our semantic communication system, we adopt an E2E learning mechanism, where our objective is to jointly achieve Eqns. \eqref{eq:cpobj}, \eqref{eq:klobj} and \eqref{eq:chobj}. Overall, we propose to minimize the following objective function $\mathcal{L}_{\xweights, \yweights}^{\alpha, \beta}$:
\begin{align}\label{eq:sloss}
    \mathcal{L}_{\xweights, \yweights}^{\alpha, \beta} {~}={~}& I_{\xweights}(X;M) - (1+\alpha) I_{\xweights}(X;Y) + \beta \mathrm{EKL}_{\xweights, \yweights}, 
\end{align}
where $\alpha \geq 0$ and $\beta \geq 0$ are some hyperparameters that trade-off the optimization. To minimize $\mathcal{L}_{\xweights, \yweights}^{\alpha, \beta}$, we hinge on the well-known cross-entropy (CE) loss defined as:
\begin{align}\label{eq:celoss}
    \mathcal{L}_{\xweights, \yweights}^{\rm CE} &\overset{\Delta}{=} \mathbb{E}_{m\sim p_{\mathcal{M}}(m),y\sim p_{\xweights}(y|m) }\left\{-\mathrm{log}(q_{\yweights}(m|y)\right\}.
\end{align}

Indeed, we have the following Lemmas.
\begin{lemma}\label{lem:lem1}
    Assuming the RX and the TX are sharing the same background \ien, $\mathcal{M}_t=\mathcal{M}_r = \mathcal{M}$, the cross-entropy loss can be decomposed as follows:
    \begin{align} \label{eq:lem1}
        \mathcal{L}_{\xweights, \yweights}^{\rm CE} = H_{\xweights}(X) - I_{\xweights}(X;Y) + \mathrm{EKL}_{\xweights, \yweights}.
    \end{align}
    \begin{proof}[Sketch of proof] First, consider the definition of $\mathrm{EKL}_{\xweights, \yweights}$ in Eq. \eqref{eq:klfunc}. Also, we have by definition, \[\mathcal{L}_{\xweights, \yweights}^{\rm CE}
    \overset{\Delta}{=} -\sum_{m\in\mathcal{M}} p_{\mathcal{M}}(m) \int_y p_{\xweights}(y|m) \mathrm{log}(q_{\yweights}(m|y)) \dy.\]
    Hence, noting that $p_{\xweights}(y|m) = \int_x p(y|x) \delta(x-f_{\xweights}(m)) \dx$, the proof follows by jointly applying Eqn. \eqref{eq:probx} with Bayes Theorem. Full proof is omitted due to lack of space.
    \end{proof}
\end{lemma}

\begin{lemma}\label{eq:lem2}
    If $\alpha\geq0$ and $0\leq\beta\leq1$, then, the objective function \eqref{eq:sloss} admits an upper-bound as follows:
    \begin{align}
        \mathcal{L}_{\xweights, \yweights}^{\alpha, \beta} \leq \mathcal{L}_{\xweights, \yweights}^{\rm CE} - \alpha I_{\xweights}(X;Y).
    \end{align}
    In particular, equality holds if $\beta=1$ and if there is no semantic redundancy at the source.
    \begin{proof}
        The proof follows by noting that $\mathcal{L}_{\xweights, \yweights}^{\alpha=0, \beta=1} = \mathcal{L}_{\xweights, \yweights}^{\rm CE} - H_{\xweights}(X|M)$ and that $H_{\xweights}(X|M)\geq 0$.
    \end{proof}
\end{lemma}

Thus, to minimize $\mathcal{L}_{\xweights, \yweights}^{\alpha, \beta}$, we can minimize this upper-bound, where $I_{\xweights}(X;Y)$ can be estimated using mutual information neural estimator \cite{MINE18}.

\begin{remark}
    Note that in \cite{Huiqiang2021a}, the authors have considered minimizing $\mathcal{L}_{\xweights, \yweights}^{\rm CE} - \alpha I_{\xweights}(X;Y)$,
    where $0\leq \alpha\leq 1$. However, the paper fails in providing a justification on how the proposed loss optimizes the semantic representation learning. In contrast, our Lemma \eqref{eq:lem2} specifies that the \emph{semantic representation loss} \eqref{eq:sloss} admits $\mathcal{L}_{\xweights, \yweights}^{\rm CE} - \alpha I_{\xweights}(X;Y)$ as an upper-bounded. Hence, minimizing this upper bound also minimizes $\mathcal{L}_{\xweights, \yweights}^{\alpha, \beta}$.
\end{remark}

\section{Transformers enabled semantic communications} \label{sec:E2E-sys}
\begin{figure*}[!t]
    \centering
    \includegraphics[width=\textwidth]{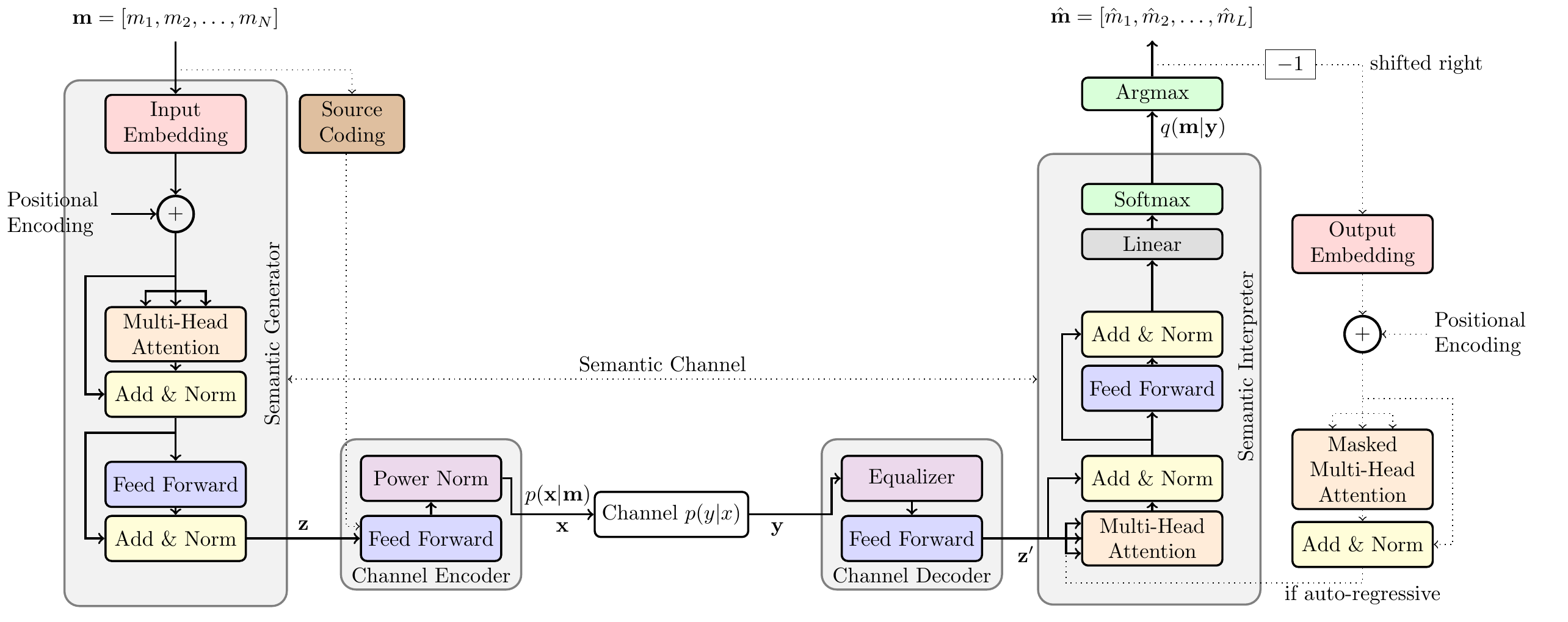}
    \caption{Transformer-based semantic communication system architecture.}
    \label{fig:sys-model}
\end{figure*}
Our semantic communication system relies on Transformer architecture \cite{vaswani2017attention}. Transformer networks have been introduced as the first transduction model entirely built using self-attention mechanisms able to learn context representation of its input and output. In contrast to solutions based on recurrent and convolution neural networks, Transformer models in general have i) lower computational complexity, ii) more parallelizable computations,  and iii) can learn long-range dependencies in input sequence \cite{vaswani2017attention}. The key components of Transformers are \emph{multi-head attention} mechanisms \cite{vaswani2017attention}. The fundamental idea behind multi-head attention is that each attention head, through its projectors, can extract specific characteristics of inputs sequences. Doing so allows the model to jointly attend to information from different representation subspaces at different positions. This aspect of multi-head attention mechanisms makes them particularly suitable for semantic information extraction, especially for NLP. Due to lack of space, we refer readers to work \cite{vaswani2017attention} for full description.

\subsection{Architecture description}
\fig{fig:sys-model} shows our proposed E2E semantic communication system composed of a source coder $S(\cdot)$, a semantic generator $G(\cdot)$, a channel encoder $E(\cdot)$, a channel decoder $D(\cdot)$, and a semantic interpreter $I(\cdot)$. 

\vspace{0.1cm}
\noindent
\textbf{Semantic generator.} %
The key component of the semantic generator is multi-head attention block (see \fig{fig:sys-model}). It allows features extraction and to find intrinsic relationships between pair of messages ($m_i$, $m_j$) in an input sequence $\mb=[m_1, m_2, \dots, m_N]$ generated by the source, where $m_i\in \mathcal{M}_t$. It outputs $\zb= G(\mb) \in \mathbb{R}^{N\times M}$, in the semantic representation subspace, mapping each message $m_i$ into $\mathbb{R}^{M}$.

\vspace{0.1cm}
\noindent
\textbf{Channel encoder.} %
It encodes the message $S(\mb)$ generated by the source coding block (\egn, using Huffman source coding), with the semantic information provided by the semantic generator $G(\mb)$: $\xb=E(\zb, S(\mb))$. Here, $E(\cdot)$ is composed of a feed-forward neural network (FNN), followed by a power normalization layer such that $\mathbb{E}[\norm{\xb}] = 1$ to average the energy of the symbols constellation. Then, each message $m_i$ is encoded in $n$ complex symbols to be transmitted through the wireless channel.

\vspace{0.1cm}
\noindent
\textbf{Wireless channel.} %
The channel outputs $\yb = h \xb + \mathrmbold{n}$, where $h$ is the fading coefficient matrix, and $\mathrmbold{n}\sim \mathcal{C}\mathcal{N}(0, \sigma_n^2\mathrmbold{I})$ is an additive Gaussian noise with power $\sigma_n^2$ and $\mathrmbold{I}$ denotes the identity matrix.

\vspace{0.1cm}
\noindent
\textbf{Channel decoder.} %
The decoder performs a channel equalization \egn, using Zero Forcing (ZF) method and decodes the received symbols into the semantic representation subspace, $\zpb=D(\yb)$ using a feed-forward neural network.

\vspace{0.1cm}
\noindent
\textbf{Semantic interpreter.} %
It plays the inverse role of the generator. it interprets the decoded semantic symbols in the space of possible messages of the receiver alphabet $\mathcal{M}_r$. As the generator, the interpreter is composed of a multi-head attention network. For each decoded message $z'_i$, the output of the interpreter is a probability distribution over all possible messages in $\mathcal{M}_r$: $[q(m| z'_i), ~\forall m \in \mathcal{M}_r]$. Each $z'_i$ is then interpreted as the message $m\in\mathcal{M}_r$ that maximizes $q(m| z'_i)$:
\begin{equation}
    \hat{m}_i = \underset{m'\in \mathcal{M}_r}{\argmax~} q(m' | z'_i), ~~\forall i.
\end{equation}

\begin{remark}
  Note that the semantic interpreter can also adopt an auto-regressive model, where the previously interpreted message is consumed as an additional input when interpreting the next one.  Auto-regressive models are particularly suitable when there is a strong correlation between different messages in the sequence (\egn, text translation). However, it requires interpreting symbols one after the other, resulting in a large decoding overhead.
\end{remark}

\subsection{Performance measure}
To assess the performance of our proposed E2E semantic communication system, we define the following measures:

\noindent
\textbf{Average transmission rate (bits/s).} %
Let $T_s$ denotes the transmission duration of each symbol. We define the average transmission rate $R$ as the ratio between the amount of transmitted information $I(X;Y)$ and $T_s$, \ien,
\begin{align}\label{eq:metric_rate}
    R = \frac{I_{\xweights}(X;Y)}{T_s}~~\text{(bits/s)}.
\end{align}

\noindent
\textbf{Accuracy vs. complexity trade-off.} %
Moreover, we also consider the following metric:
\begin{align}\label{eq:metric_tau}
    \tau = \frac{1}{\mathbb{E}[n]} \times (1 -\psi_{\xweights, \yweights}(M, \hat{M})),
\end{align}
where $\mathbb{E}[n]$ is the average number of symbol per transmitted message. Here, $\psi_{\xweights, \yweights}(M, \hat{M})$ measures the semantic error between transmitted message $M$ and interpreted message $\hat{M}$. This error takes different forms depending on the context \cite{kountouris2020} (\egn, mean square error, cross-entropy or BLEU score in NLP \cite{bleu2002}). Thus, $\tau$ measures the trade-off between ``transmission accuracy" and model complexity in terms of average number of symbols ($\mathbb{E}[n]$) used to represent each message.

\section{Numerical Results} \label{sec:semantic-results}

We provide a detailed evaluation of the performance of our proposed adaptive E2E semantic communication system in the context of natural language processing. Numerical results are reported for text transmission as in \cite{Huiqiang2021a}. Our reference scenario considers a transmitter communicating with a receiver by sending a block of sentences (sequence of words) through the wireless channel using the previously described semantic communication system. To this end, the transmitter learns to map each word to a sequence of semantic symbols that the receiver has to interpret. Note that such a mapping is learned from the data available at the source. Hence a word can have different symbols representation depending on the sentence it belongs to and the underlying meaning conveyed by both the word and the sentence. This is in contrast to traditional Level A communication, where each word is always mapped to the same symbol. In our scenario, once received symbols are interpreted back to words, we measure the transmission accuracy in terms of bilingual evaluation understudy (BLEU) score, which counts the difference of words (or group of words - n-grams) between the intended sentence and interpreted one \cite{bleu2002}. Its value range from zero to one, with one indicating that the interpreted message is the one as the reference. We consider averaging the BLEU score over 1-gram to 4-grams. We use the dataset from Tatoeba Project (translation from English to French data available at http://www.manythings.org/anki/).
All FNNs are composed of one multi-layer perceptron with 128 neurons and we use $6$ attention heads. Unless otherwise specified, we set $M=64$, $T_s=1s$ (normalized), $n=6$, $\alpha=0.01$ and $\beta=1$. Then we train the proposed E2E network for a reference signal-to-noise ratio (SNR) of $7$ dB using a batch-size of 256 and then performs tests for different SNRs.

\begin{figure}
    \centering
    \includegraphics[width=0.81\columnwidth]{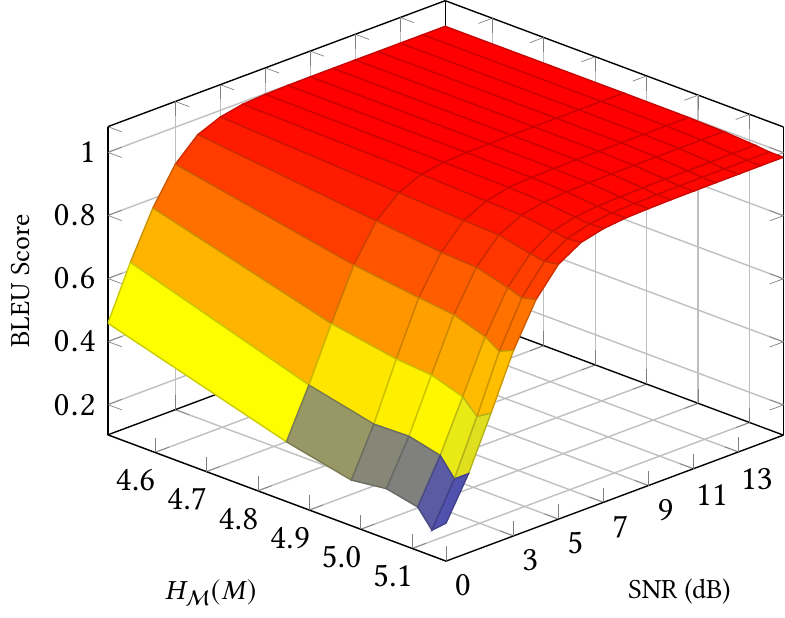}
    \caption{Impact of the SNR and $H_\mathcal{M}(M)$ on the accuracy. Here we use $n=6$ symbols/word over AWGN channel.}
    \label{fig:vocab-impact}
\end{figure}

\vspace{0.15cm}
\noindent
\textbf{Impact of the SNR and the source entropy.} %
We first show in \fig{fig:vocab-impact}, the impact of SNR and the source entropy on transmission accuracy. We change the source entropy by modifying the distribution $p_{\mathcal{M}}(M)$. We observe in \fig{fig:vocab-impact} that the performance slightly decreases when $H_\mathcal{M}(M)$ increases since there is more information to convey to the receiver. Also, we observe that the proposed scheme achieves a BLEU score of 1 for $\mathrm{SNR}\geq5 ~\mathrm{dB}$. In particular, we observe that this threshold varies with the reference SNR for the training, which we set here to $7~\mathrm{dB}$.

\begin{figure}[!t]
    \centering
    \includegraphics[width=0.95\columnwidth]{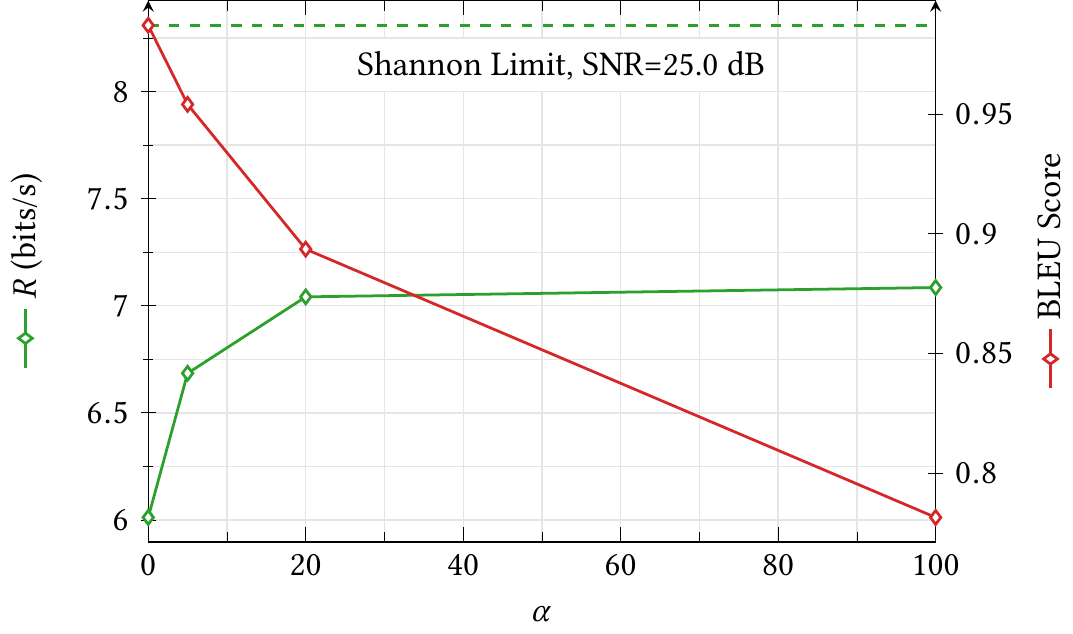}
    \caption{Impact of the trade-off parameter $\alpha$ on performances. }
    \label{fig:impact-alpha}
\end{figure}

\vspace{0.1cm}
\noindent
\textbf{Impact of $\alpha$.} %
Here, we assess the performance of the proposed scheme \wrt the trade-off parameter $\alpha$ of the objective function \eqref{eq:sloss}. \fig{fig:impact-alpha} shows the BLEU score and the mutual information of the channel for different $\alpha$. As $\alpha$ increases we give more importance to $I(X;Y)$ term \eqref{eq:sloss}, thus increasing the mutual information at the risk of degrading the accuracy.

\begin{figure}[!t]
    \centering
    \includegraphics[width=0.95\columnwidth]{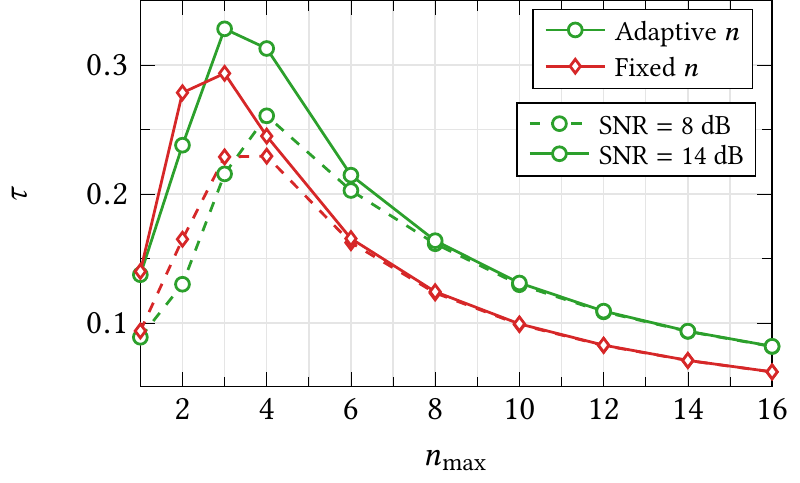}
    \caption{Adaptive vs Fixed number of symbols/word.}
    \label{fig:adaptive-encoding}
\end{figure}

\vspace{0.1cm}
\noindent
\textbf{Impact of the number of symbols per word.} %
Authors in \cite{Huiqiang2021a} consider a fixed number of symbols per word sent through the channel. However, depending on the length of the words (\egn, the number of characters) and/or the conveyed semantic information, different words may not use the same number of symbols. To show this effect, let $\mb=[m_1, m_2, \dots, m_N]$ be a sequence of words to be transmitted and $l(m_i)$ the length of each word $m_i$ on a character basis. Let $L_{\mb} = \sum_i l(m_i)$ be the total number of characters in sequence $\mb$. we first construct the probability vector $\mathrmbold{p}=[p_1, \dots, p_N]$ such that $p_i = \frac{l(m_i)}{L_{\mb}}, ~~\forall i$. Hence, $p_i$ defines the weight of the word $m_i$ in the sequence in terms of number of characters. Now, let $n_{\max}$ be the maximum number of symbols admissible for each word. Then, we encode each word $m_i$ in $n_i$ (instead of fixed $n = n_{\max}$ as considered in \cite{Huiqiang2021a}) symbols 
where,
\begin{align}
    n_i = \min\left(\max\left(n_{\min}, \lfloor n_{\max}N p_i + \frac{1}{2}\rfloor\right), n_{\max}\right).
\end{align}
Hence, $n_i\in [n_{\min},n_{\max}], ~\forall i$. In \fig{fig:adaptive-encoding}, we show the impact of the adaptive vs. fixed encoding on the metric $\tau$, where we arbitrary fix $n_{\min}=1$ and let $n_{\max}\in[1,16]$. We first note that in both cases, there is a trade-off between accuracy and complexity, \ien, there is an optimal value ($n^{*}$) of $n_{\max}$ depending on the SNR. In particular, for the fixed case ($n=n_{\max}$), and for lower SNR ($8$dB) we have $n^{*}=4$. As the SNR increases to $14$dB, only $n^{*}=3$ symbols are sufficient to encode each word. In the adaptive case, the number of symbols per word is adapted to the words' length such that on average, $\mathbb{E}[n] \leq n_{\max}$. Therefore, in \fig{fig:adaptive-encoding}, we clearly see that when $n_{\max}\leq4$, the adaptive method outperforms the fixed one, exhibiting $21.7\%$ increase in $\tau$. This means that for the same accuracy, the adaptive encoding uses a lower number of symbols than the fixed encoding to represent each word. When $n_{\max}\leq3$, the adaptive method is slightly less efficient: this suggests that there is a minimum number of symbols per word to meet a given accuracy, here \egn, $n_{\min}=2$.
\begin{figure}[!t]
    \centering
    \includegraphics[width=0.95\columnwidth]{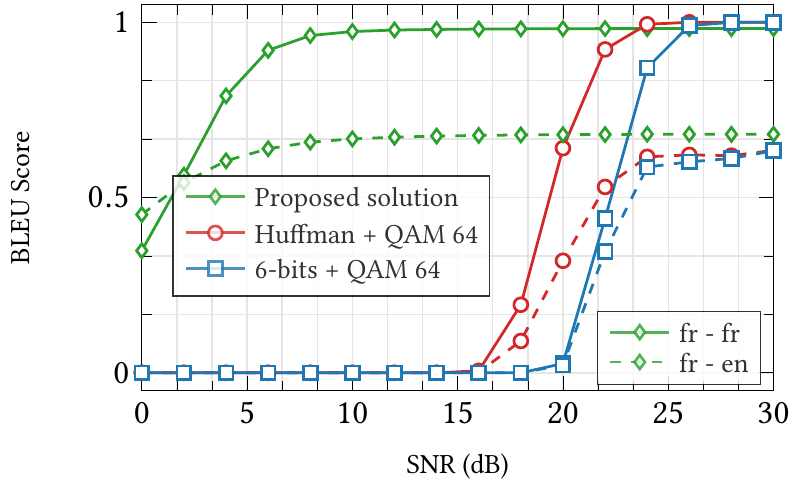}
    \caption{1-gran BLEU Score vs. SNR in the context of AWGN channel for French-to-(French/English) translation.}
    \label{fig:translation}
\end{figure}

\vspace{0.1cm}
\noindent
\textbf{Impact of languages mismatch.} %
We now show a scenario where the transmitter speaks French and the receiver must understand in English. In this case, the sender and the receiver have different alphabets. This further introduces complexity in symbols interpretation. Indeed, many words in French are written the same way in English leading to semantic ambiguity. The result is $30\%$ decrease in BLEU score performance as show in \fig{fig:translation}. In the same figure, we also show the performance of the classical approach using Huffman/6-bits coding and a 64 QAM modulation. Note that as there is no way to infer English words from decoded symbols in the classical approaches, we rely on Google Translator, although its alphabet is larger than that of our receiver. The proposed semantic communication clearly outperforms the two benchmarks, especially in the low SNR regime.



\section{Conclusion}

In this work, we focused on the benefit offered by semantic compression to beyond 5G communications. To this end, we proposed a novel E2E architecture for an efficient semantic communication system. We started by analyzing theoretical aspects to formulate an objective function for semantic representation learning. Then, we proposed a new metric and trade-off parameter to assess the performance of the proposed system in terms of transmission accuracy and model complexity. Eventually, we provided an example on text transmission, which shows a significant semantic compression gain, especially when sender and receiver speak different languages. In this example, the sender learns to map transmitted sentences into a sequence of well-formed symbols, exploiting the semantic, \ien, the meaning conveyed by these sentences. Then, we proposed a mechanism that adapts the number of symbols per word based on the conveyed semantic, providing up to $21\%$ extra gain compared to state-of-the-art approaches. Importantly, this gain can be significantly extended when applied to multi-modal and data-angry applications such as video-to-text or text-to-video.

\vspace{0.35cm}
\bibliographystyle{IEEEtran}
\bibliography{biblio}

\end{document}